\documentclass[prl,aps,twocolumn,floats,nofootinbib,superscriptaddress]{revtex4-1}
\usepackage{graphicx}
\usepackage{dcolumn}
\usepackage{bm}
\usepackage[colorlinks=true,urlcolor=blue,citecolor=blue]{hyperref}
\usepackage{float}
\usepackage{subfig}

\usepackage{amsmath}
\usepackage{amssymb}
\usepackage{float}
\usepackage{physics}
\usepackage{comment}

\begin{document}


\title{Periodic Drive Induced Half-Metallic Phase in Insulators and Correlated Metals}

\author{Suryashekhar Kusari}
 \affiliation{Theory Division, Saha Institute of Nuclear Physics, 1/AF Bidhannagar, Kolkata 700064, India}
\affiliation{Homi Bhabha National Institute, Training School Complex, Anushaktinagar, Mumbai 400094, India}
\author{Arnab Das}
\affiliation{Indian Association for the Cultivation of Science (School of Physical Sciences), P.O. Jadavpur University, 2A \& 2B Raja S. C. Mullick Road, Kolkata 700032, West Bengal, India}
\author{H. R. Krishnamurthy}
\affiliation{Center For Condensed Matter Theory, Department of Physics, Indian Institute of Science, Bangalore 560012, India}
\affiliation{International Centre for Theoretical Sciences, Tata Institute of Fundamental Research, Bengaluru 560089, India}
\author{Arti Garg}%
 \affiliation{Theory Division, Saha Institute of Nuclear Physics, 1/AF Bidhannagar, Kolkata 700064, India}
\affiliation{Homi Bhabha National Institute, Training School Complex, Anushaktinagar, Mumbai 400094, India}

\begin{abstract}
  Non-equilibrium control of electronic properties in condensed matter systems can result in novel phenomena. In this work, we provide a novel non-equilibrium route to realize half-metallic phases. We explore the periodically driven Hubbard model on a bipartite lattice and demonstrate that a specially designed sublattice dependent drive can transform a weakly interacting metal or insulator into a ferrimagnetic half-metal (HM).  We consider a Fermi-Hubbard model with only nearest-neighbour hopping and Floquet engineer the elusive half-metal phase by driving the site potentials periodically. The drive induces staggered higher range hoppings and a staggered potential between two sublattices in the Floquet Hamiltonian. Close to the dynamical localization point, due to the suppression of nearest neighbor hopping in the driven system, an effective enhancement of various terms in the Floquet Hamiltonian, including the e-e interactions, occurs. This helps in stabilizing a broad ferrimagnetic HM phase for a wide range of drive parameters. This dynamically stabilized HM phase is a pristine platform for next-generation spintronics, topological quantum computing (specifically for engineering Majorana zero modes in hybrid nanowires) and ultrafast magnetic memory architectures.
  
\end{abstract}

\maketitle

\section{Introduction}
\label{sec-1}
\noindent
Periodically driven quantum systems are known to host interesting phases that are otherwise not present in the system in equilibrium. Some of the most remarkable examples of this are: unconventional superconductivity induced by exciting samples with lasers in the non-superconducting phase~\cite{light_sc_cuprates,light_cuprates2,fulleren_sc}, light-induced anomalous Hall effect in monolayer graphene~\cite{graphene_AQH} and on the 2D surface of a 3D topological insulator~\cite{AQHE}, light-induced many-body localization~\cite{MBL_light}, creation of long-range order like charge-density-wave and spin-density wave order~\cite{AFMorder,charge_order}, or even melting of the existing long-range order~\cite{melting} by exciting samples with high-frequency light and tuning the transition temperature in conventional superconductors~\cite{Komnik}. 
Periodically driven cold atomic systems have recently engineered models otherwise inaccessible in static systems, e.g., drive-engineered gauge fields~\cite{gauge}, Haldane model~\cite{Haldane}, and other topological systems~\cite{Harper, Hofstadter}. In this work, we propose a novel mechanism to realize a half-metal(HM) using a periodic drive starting from a weakly correlated metal or an insulator.  

Generic periodically driven systems are expected to absorb energy indefinitely and reach a locally infinite-temperature-like state~\cite{LDM_PRE, Alessio2014} when the drive parameters are of the same order as that of the leading couplings of the undriven part of the Hamiltonian. However, two different regimes of stability are known to emerge when (a) the drive frequency $\Omega$ is large and (b) when the drive amplitude is large compared to the leading parameters of the static part of the Hamiltonian, but not so large that they cause excitations into higher energy bands. In case (a) one observes a long prethermal stability: For sufficiently large driving frequency $\Omega$, that is, $\Omega\gg W$ where $W$ is the local bandwidth of the system, these heating times become exponentially large $t_{th} \ge O(\exp{\Omega/W})$~\cite{Dima_Pretherm_PRL, Mori_Kuwahara_Saito_Prethermal_PRL, Kuwahara_Mori_Saito, Dima_Floquet_Prethermalization,
Prethermal_Without_T, Ho_Mori_Abanin_Pretherm_Rev,expt_DF}. In the other case, (b), when the drive field is strong enough 
the heating rate is further suppressed~\cite{Mori_2022}.
Additionally an interesting scenario known as dynamical freezing (DF)~\cite{expt_DF} emerges in this regime of the drive.
In the DF regime, new approximate but stable conservation laws emerge that arrest thermalization and stabilize the system to non-trivial states with controlled entanglement entropy~\cite{AD-DMF, AD-SDG, Mahesh_Freezing, Naveen_Dynamical_Freezing, Onset, Asmi_DF_PRX_2021, Asmi_Flq_Rev, Debanjan_FlowEq_DF, Debanjan_Frozonium_DF, Adhip_Diptiman, Diptiman_DF, Bhaskar_DF, Somsubhra_Paul_Krishnendu_DF,DF_TDL_iTEBD}.The stability of such clean, interacting "Floquet matter", covering and far surpassing any experimentally accessible regime, is hence well-established now. Thus, working with high frequency and high drive amplitude provides an alternative route toward engineering rich, dynamically stabilized systems hosting novel phases of matter not easily realizable in an equilibrium setting. \\

\noindent
Half-metallic states, characterized by fully spin-polarized carriers, are central to quantum technologies due to their intrinsically spin-selective transport. They provide ideal spin filters for spintronic devices and enable proximity-induced triplet pairing in superconducting hybrid structures, forming key elements of superconducting spintronics and topological architectures \cite{Zutic2004,Eschrig2015}. Dynamically stabilizing such spin-selective phases is particularly promising in programmable quantum simulators—including superconducting qubit arrays, ultracold atoms in optical lattices, and trapped ions—where Floquet engineering enables controlled realization of many-body Hamiltonians and transport phenomena \cite{Bloch2008}. Furthermore, when coupled to conventional superconductors, half-metals naturally realize topological superconductivity and Majorana zero modes due to their intrinsic spin polarization, which produces an effective spinless system supporting proximity-induced topological pairing \cite{Duckheim2011,Chung2011}. Dynamical stabilization of half-metallic phases therefore provides a powerful route for engineering robust spin-polarized transport and controllable topological quantum states across both solid-state and synthetic quantum platforms.
\\
\noindent
We demonstrate a Floquet-engineering strategy to realize a stable half-metal phase in the Fermi-Hubbard model on a bipartite lattice via a sublattice-dependent periodic drive. We show that such a drive transforms its weakly interacting metallic or insulating phase into an HM phase in which electrons with one spin polarization conduct, whereas electrons with the opposite spin polarization are insulating. The drive induces a large staggered potential that provides different environments for up and down spins making the single-particle excitation gaps different for up- and down spin channels. A doped system can then easily achieve an HM phase with a shift of the Fermi level into the valence band of only one of the spin components.
This mechanism is implementable in cold atom experiments, where driven~\cite{Esslinger1,Esslinger2} and undriven  Hubbard models ~\cite{Hubbard_Bloch,Hubbard_Insbruk, cold_atom_Hubbard,Google_Fermi_Hubbard,Hubbard_Esslinger,cold_atom_Hubbard} have already been simulated, and holds promise for future realization in artificial superlattices ~\cite{Cano,Chandni} and twisted transition metal dichalcogenide heterostructures like $MoTe_2/WSe_2$ heterobilayers~\cite{TMD1,Li}, $MoSe_2/WSe_2$ heterostructures ~\cite{TMD2} etc.(see Appendix-A).
\begin{figure}[t]
    \centering
    \vspace{-2.4cm}
    \includegraphics[width=1.0\columnwidth,height=1.0\columnwidth]{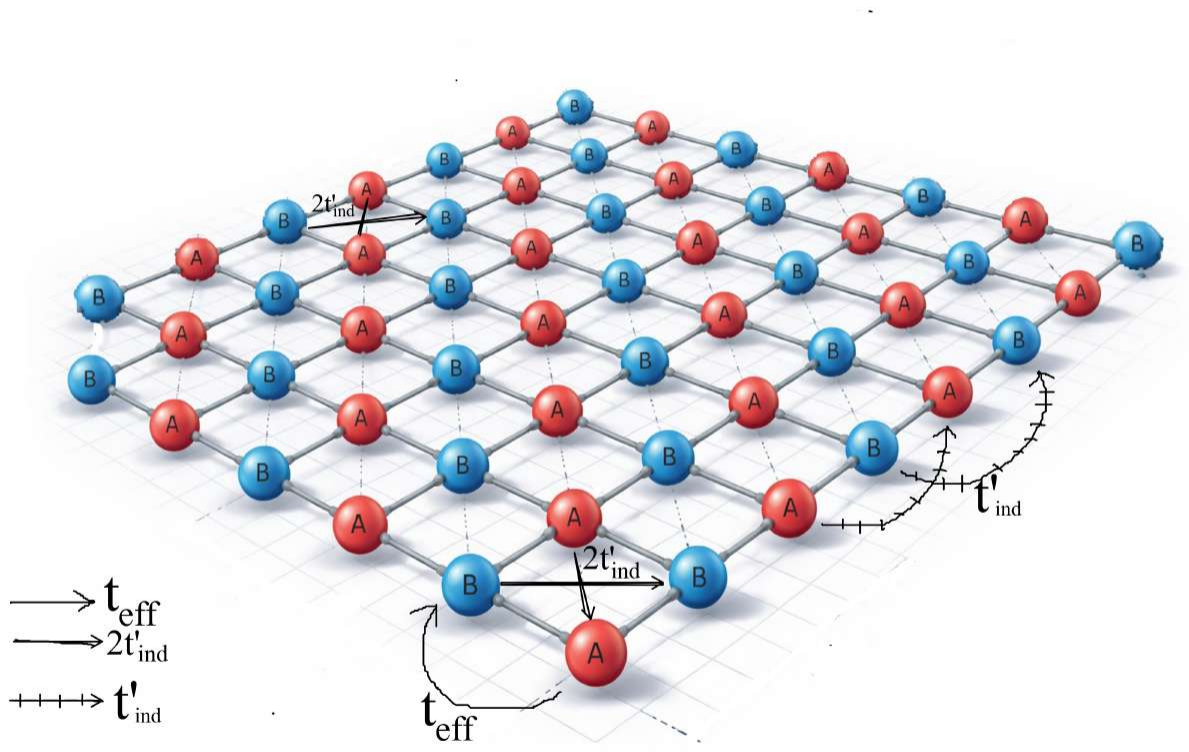}

\vspace{-3cm}
    \caption{Schematic diagram to show drive induced hoppings on a 2d-square lattice.} 
    \label{lattice}
\end{figure}

\section{Periodically Driven Hubbard Model and its Floquet Phases}
The model we consider is described by the following time-dependent Hamiltonian.     
\[H(t) =H_{Hub}+H_{drive}(t)\] 
\begin{equation*}
 H_{Hub}=-t_0\sum_{\langle ij\rangle,\sigma}c^\dagger_{iA\sigma}c_{jB\sigma}+ h.c.+U\sum_{i}n_{i\uparrow}n_{i\downarrow}-\mu\sum_i n_i 
 \end{equation*}
 \begin{equation}
 H_{drive}(t)=\frac{V}{2}\sum_{i\in \alpha}Sin\big(\Omega t-sgn(\alpha)\Phi\big)\hat{n}_{i}
\label{hamil}
\end{equation}
Here, $sgn(\alpha)=\pm$ for $\alpha=A,B$, $t_0$ is the nearest-neighbor hopping (indicated by $<ij>$), $U$ is the onsite repulsion between electrons, and $\mu$ is the chemical potential that tunes the average density in the system. $H_{drive}(t)$ describes a time-periodic single particle Hamiltonian with driving frequency $\Omega$ and amplitude $V$, which are the same for both sublattices, and a phase difference $2\Phi$ between the two sublattices. In cold-atom experiments independent driving of sublattices A and B can be achieved by attaching piezoelectric actuators to the retro-reflecting mirrors of two optical standing waves oriented at $45^\circ$ and $135^\circ$. The inter-sublattice drive phase shift $\Phi$ is introduced by turning on the drives with a time lag $\Delta t = 2\Phi/\Omega$ (Appendix-A).

The time dependence in the above Hamiltonian can be transformed into an effective time-dependent hopping by going to a rotating frame $H_{rot} = S^\dagger(t) H (t) S(t)-iS^\dagger(t)\partial_t S(t)$, where $S(t)=\exp\big(-i\sum_j F_j(t) n_j\big)$ and $F_j(t)=\frac{V}{2}\int_0^t \sin(\Omega t \mp \Phi)$ for $j\in A(B)$ respectively~\cite{Polkovnikov,Eckardt_rev}.  In the rotating frame, the kinetic energy is
$H_{hop}(t)=-t_0g(t)\sum_{\langle ij\rangle,\sigma}c^\dagger_{iA\sigma}c_{jB\sigma}+ h.c.$ with $g(t)=\exp(i\frac{V}{\Omega}\sin(\Omega t)\sin(\Phi))$, while $H_{drive}$ is eliminated. The other terms in the Hamiltonian $H_{hub}$ remain unchanged under rotation. 

\begin{figure}[t]
    \centering
    \includegraphics[width=1.0\columnwidth,height=0.7\columnwidth]{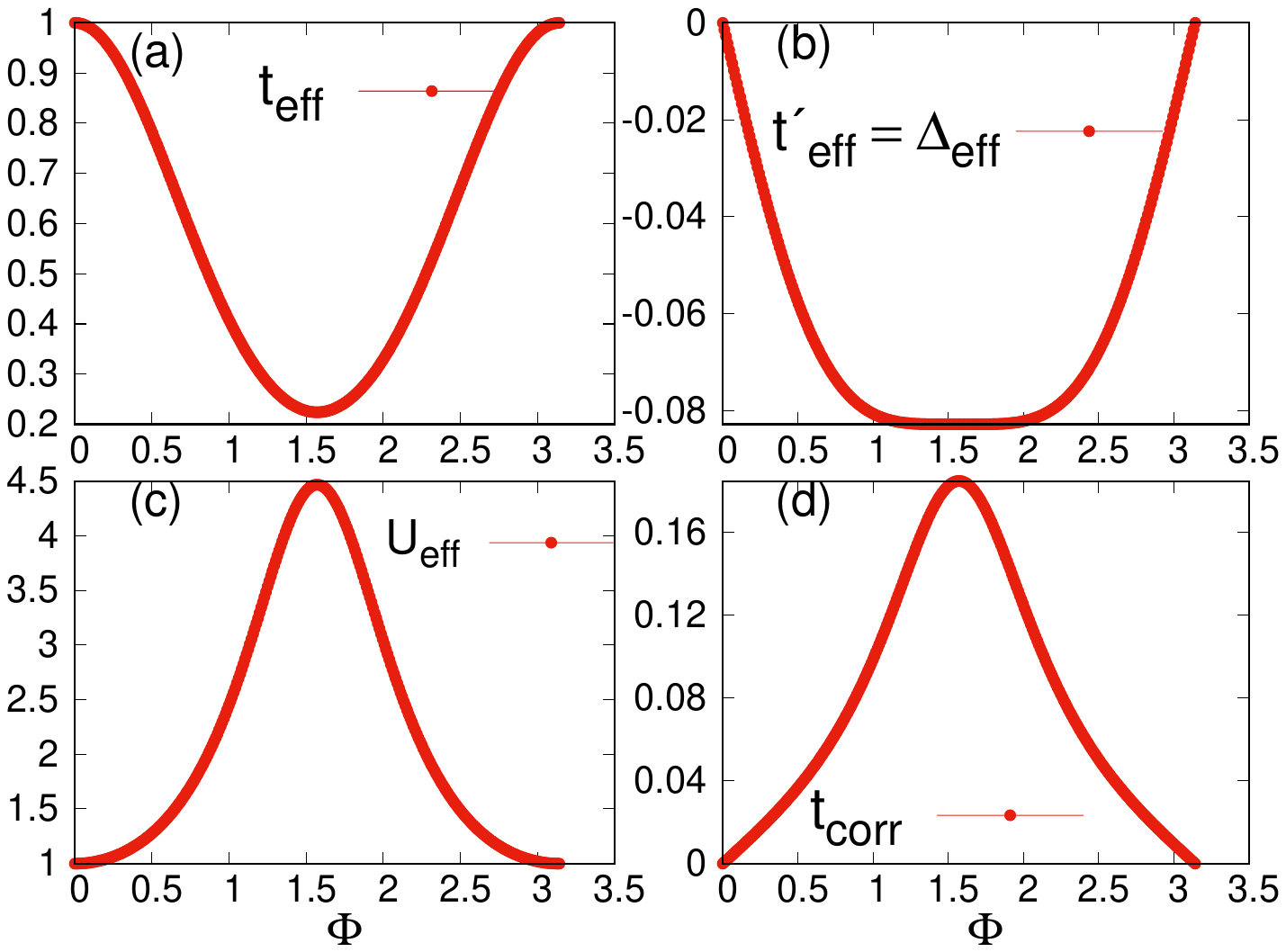}
    \caption{Effective couplings as functions of $\Phi$ in the periodically driven Hubbard model for the drive with amplitude $V=60$ and frequency $\Omega=30$. $U=1.0$ is used for the couplings in the bottom panels.} 
    \label{couplings}
\end{figure}

The leading corrections to the stroboscopic Floquet Hamiltonian resulting from this periodic driving are then found from the Magnus expansion~\cite{Polkovnikov}. We will keep terms up to order $1/\Omega$ in the Magnus expansion, which results in the following Floquet Hamiltonian $H_F=H_F^{[0]}+H_F^{[1]}$ with (for details see Appendix-B).

\begin{equation*}
    H_F^{[0]} =-t_{eff}\sum_{\langle ij \rangle, \sigma} c_{i,\sigma}^\dagger c_{j,\sigma} + h.c + U \sum_j n_{j \uparrow} n_{j \downarrow} -\mu\sum_i n_i,
    \end{equation*}
\begin{equation*}
H_F^{[1]}=2\Delta_{ind}\sum_{\langle ij\rangle,\sigma}  \big( n_{i \sigma}^A - n_{j, \sigma}^B \big)+2t_{ind}^\prime\sum_{(ij),\alpha\sigma}sgn(\alpha)c_{i\alpha,\sigma}^\dagger c_{j\alpha,\sigma}
\end{equation*}
\begin{equation}
\begin{split}
+t_{ind}^{c}\sum_{\langle ij \rangle,\sigma} \big(n_{i \sigma}^A -n_{j, \sigma}^B \big)c_{iA \bar{\sigma}}^\dagger c_{jB, \bar{\sigma}} + h.c.\\
+t^{\prime}_{ind} \sum_{((ij)),\alpha\sigma} sgn(\alpha)c_{i\alpha,\sigma}^\dagger c_{j\alpha,\sigma}
\end{split}    
\label{HF1}
\end{equation}
\\
\begin{figure}[t]
    \centering
    \hspace{-1cm}
    \includegraphics[width=0.7\columnwidth,height=0.5\columnwidth]{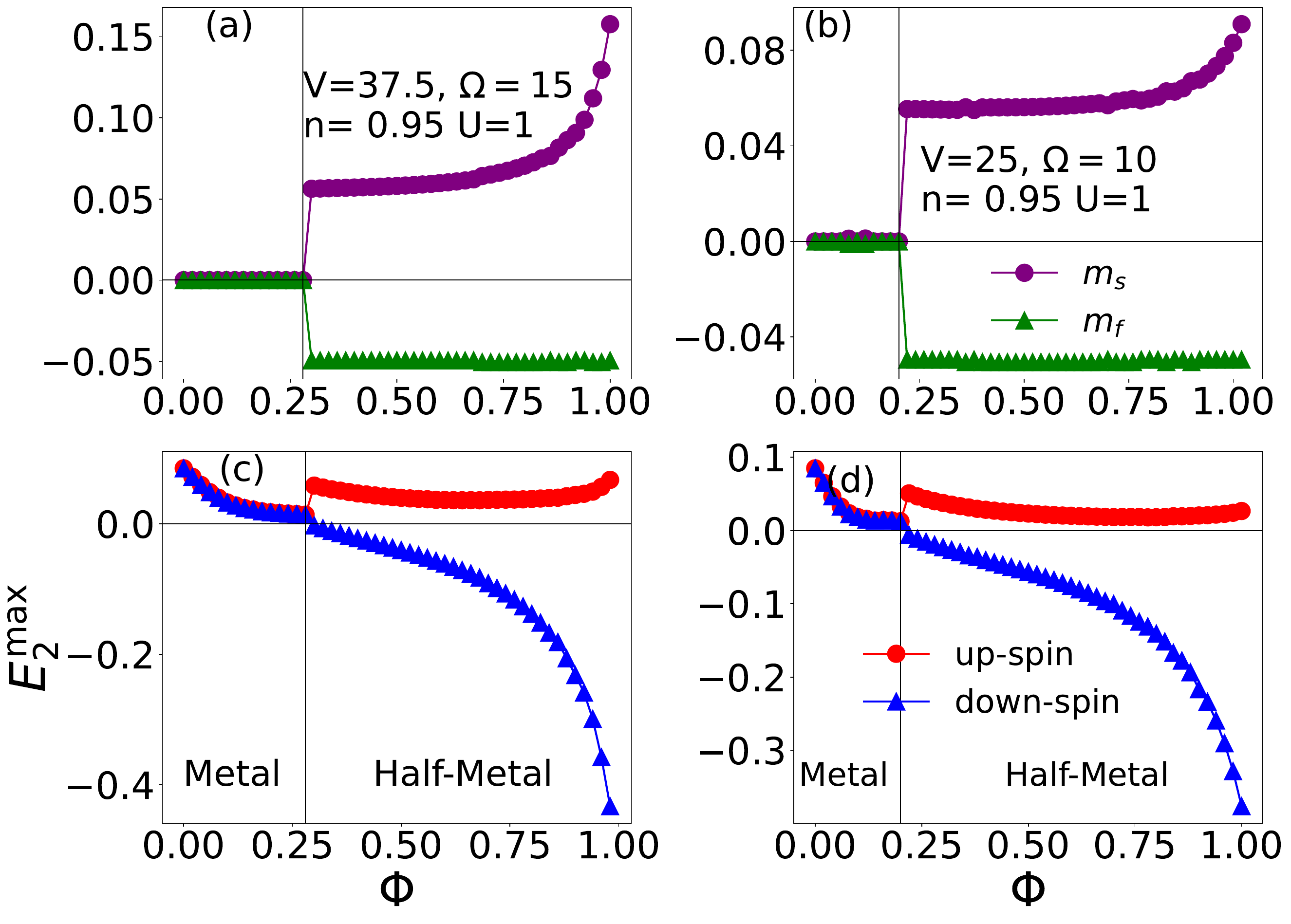}
\hspace{-0.2cm}
\includegraphics[width=0.25\columnwidth,height=0.5\columnwidth]{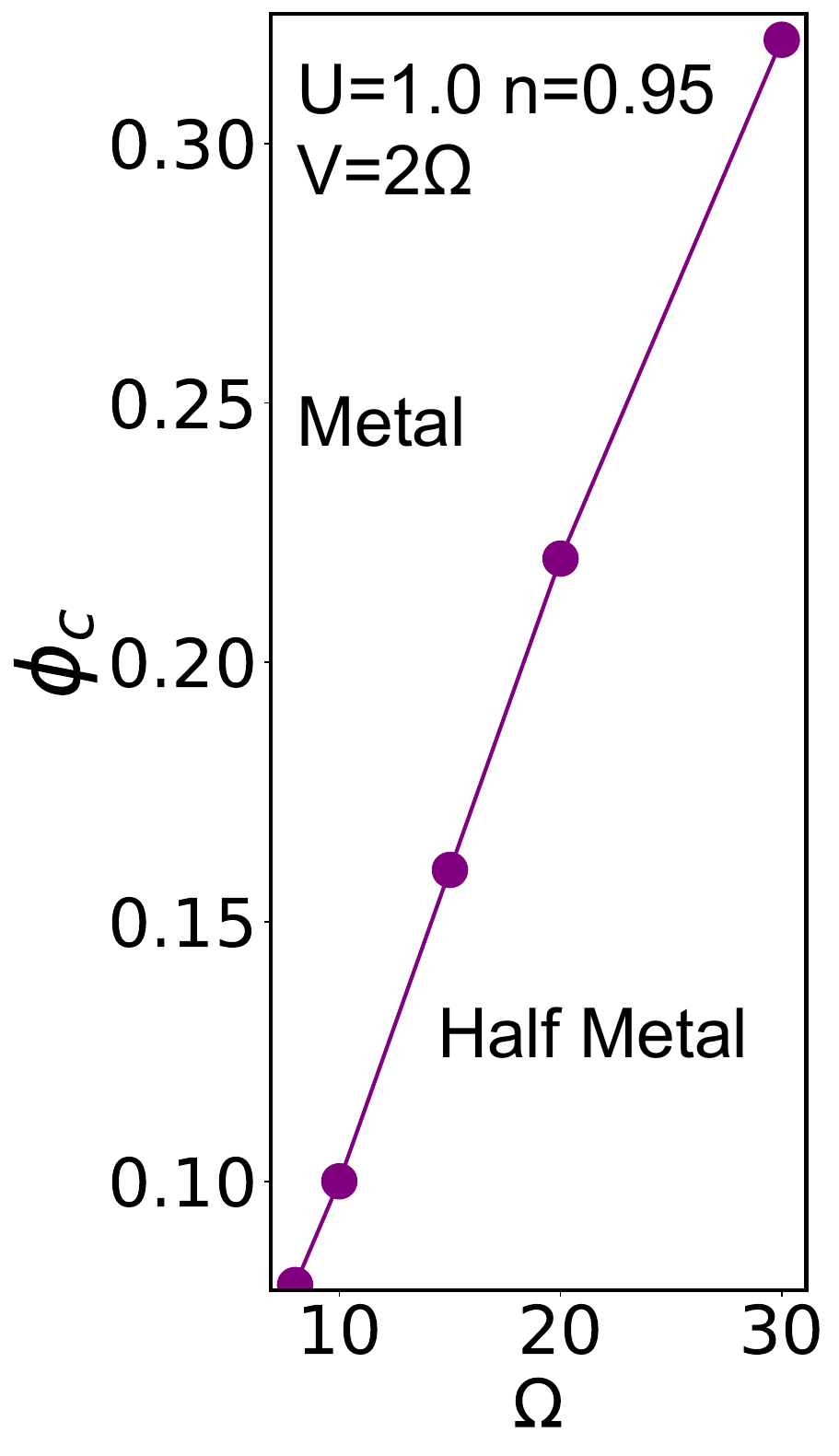}
    \caption{Panels [a,b]: Staggered magnetization $m_s$ and uniform magnetization $m_f$ vs $\Phi$ for $n=0.95$ $\Omega=15$ and $10$ with $V=2\Omega$. Panels [c,d]: $E_{2\sigma}^{max}$ as functions of $\Phi$. At $\Phi_c$, the paramagnetic metal turns into a ferrimagnetic HM. The rightmost panel shows the phase diagram in the $\Phi-\Omega$ plane for $V=2\Omega$, $n=0.95,$ and $U=1.0$.} 
    \label{ms_n}
\end{figure}
\noindent
Here, $t_{eff} = t_0 \mathcal{J}_0(\frac{V}{\Omega}sin(\Phi))$ where $\mathcal{J}_0$ is the zeroth-order Bessel function of the first kind. For the range of parameters where $\mathcal{J}_0$ hits zero, the effective hopping of nearest neighbors is suppressed, resulting in the well-known dynamical localization (DL) of the system\cite{Dunlap, Hanggi,Holthaus, Eckardt, Tsuji} which is the same as the DF for non-interacting systems~\cite{AD-DMF}. Even more interestingly, the drive induces a staggered potential $\Delta_{ind}$ between sublattices A and B that breaks the translational symmetry of the original lattice. This folds the Brillouin zone, resulting in a spectral gap in the single-particle excitation spectrum and a density difference between the two sublattices. The drive also induces second and third neighbor hoppings (indicated by $(ij)$ and $((ij))$, of strength $2t^\prime_{ind}$ and $t^\prime_{ind}$ respectively, providing dynamics within each sublattice. The various couplings are depicted in the Fig.[\ref{lattice}]  for a 2d square lattice on which we perform all our calculations. Note that higher-range hoppings induced by the drive are different in nature from conventional higher-range hopping terms. The drive-induced hopping amplitude is staggered on the two sublattices and does not break particle-hole symmetry, unlike the conventional second-neighbor hopping term. The drive also induces a correlated hopping term $t^c_{ind}$ that provides spin-dependent nearest-neighbor hopping and effectively prohibits the hopping of a doublon or a holon on the nearest-neighbor bond. Detailed expressions for $\Delta_{ind}$, $t^\prime_{ind}$, and $t^{c}_{ind}$ are provided in Appendix-B. Although the Floquet Hamiltonian has similarity to the ionic Hubbard model (IHM)~\cite{IHM1,IHM2,cold_atom_IHM} explored in several earlier works~\cite{IHM_AG1,IHM_Craco,IHM_Kampf,IHM_AG2,IHM_Hoang,IHM_Bag}, it has interesting differences from the IHM in terms of the staggered higher-range hoppings and the spin-dependent correlated hopping.

In the non-interacting limit, $U=0$, the dispersion of $H_F/t_{eff}$ is given by $E_{1,2}(k)= -\mu_{eff} \pm \sqrt{(\epsilon^\prime(k)+\Delta_{eff})^2 +\epsilon_0(k)^2}$. Here, $\epsilon_0(k)=2[cosk_x+cosk_y]$, $\epsilon^\prime(k) = -8t^\prime_{eff} cosk_x cosk_y -2t^\prime_{eff}[cos(2k_x)+cos(2k_y)]$, $t^\prime_{eff} \equiv
t^\prime_{ind}/t_{eff}$, $\mu_{eff} \equiv \mu/t_{eff}$ and $\Delta_{eff} \equiv \Delta_{ind}/t_{eff}$. 
Starting from a metallic half-filled system (for which $\mu=0$), the drive induces a spectral gap in the single-particle excitation spectrum, $E_{gap} = 2|\Delta_{eff}+4t^\prime_{eff}|$.  Thus, the staggered 2nd and 3rd neighbor hoppings add to the spectral gap along with the staggered potential, unlike a conventional second-order hopping, which can suppress the gap induced by the staggered potential. Before the analysis of the interacting system, we first discuss how various couplings in the Hamiltonian of Eqn.[\ref{HF1}] evolve as a function of the drive.\\
\begin{figure}[t]
    \centering
    \vspace{-0.55cm}
     \includegraphics[width=1.0\columnwidth,height=0.64\columnwidth]{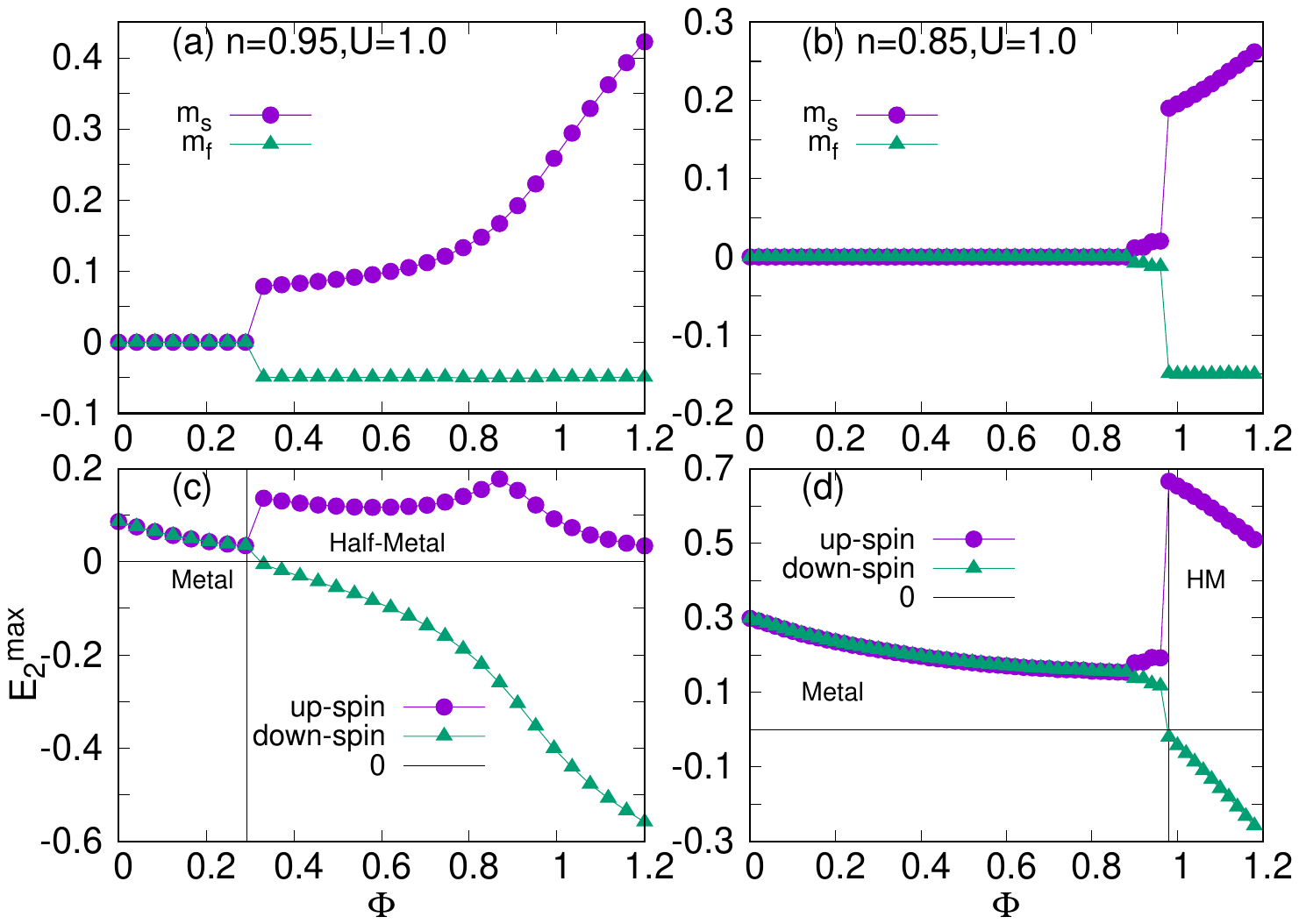}
     \vskip-0.75cm
    \caption{Top panels:$m_s$ and $m_f$ vs $\Phi$ for two different densities $n=0.95$ and $n=0.85$. Bottom panels: $E_{2\sigma}^{max}$ as functions of $\Phi$. At $\Phi_c$, a paramagnetic metal turns into a ferrimagnetic HM.
    The data shown are for $V=60$ and $\Omega=30$.}
    \label{ms_n2}
\end{figure}

\noindent
{\bf{Effective Couplings in the Driven Model :}}
Although we set the drive parameters, $V$ and $\Omega$,
slightly away from the DL point, the system still has enough suppression of $t_{eff}$. This won't affect the heating rate of the system as long as the drive frequency and amplitude are much larger than $U$ and $t_0$. Additionally, the system should exhibit dynamical freezing close to the DL point, which will be explored in future work.

The suppression in $t_{eff}$ effectively increases the e-e interactions along with a sufficient enhancement of higher range hoppings and the staggered potential in the system. This is depicted in Fig. [\ref{couplings}], which presents various couplings in $H_F$ as functions $\Phi$ for fixed values of $V=60$ and $\Omega=30$. As $\Phi$ is tuned from $0$ to $\pi/2$, the nearest-neighbour hopping $t_{eff}$ decreases, reaching its minimum at $\Phi=\pi/2$. 
In contrast, $t^\prime_{eff}$, which is also the same as the drive-induced staggered potential $\Delta_{eff}$, as well as $U_{eff}=U/t_{eff}$ and $t_{corr}=t^c_{ind}/t_{eff}$ increase as $\Phi$ increases.

\noindent
{\bf{The Ground State of $H_F$:}}
Next, we explore the ground state of the Hamiltonian $H_F$ of Eqn.~\ref{HF1} on a 2-dimensional square lattice. Of course, the concept of a ground state in its true sense of minimum energy does not apply to a Floquet system, because the quasi-energies, unlike true energies, are non-unique (by addition/subtraction of any integer multiple of the photon energy $\hbar\Omega$), being defined to correspond to a unique evolution operator rather than a unique Hamiltonian~\cite{Shirley_Floquet, Sambe_Floquet}. However, identifying the ground state of the operator $H_F,$ treating it just like a static Hamiltonian, helps one to identify the Floquet states which harbor lower quantum fluctuations against a possible local ordering.
For studying the ground state of $H_F,$ here
we employ the spin-resolved Hartree-Fock (HF) theory. The Green's function matrix within the HF theory is given by
\[
\hat{G_\sigma}^{-1}(k,\omega)= 
\begin{pmatrix}
    \xi_{A\sigma}(k,\omega) & -\epsilon_{k\sigma} \\
    -\epsilon_{k\sigma} & \xi_{B,\sigma}(k,\omega)
\end{pmatrix} 
\]
with $\xi_{A(B),\sigma}(k,\omega)=\omega^{+}\mp (\Delta_{eff}+t^\prime_{eff}\epsilon^\prime(k))+\mu_{eff}-\Sigma_{A(B),\sigma}(\omega)\equiv \omega^{+}-f_{A(B),\sigma}(k)$ where we have assumed that $\hat{G}_\sigma ^{-1}$ , $\omega$, etc., are being measured in units of $t_{eff}$. Within HF theory, the Dyson self-energy $\Sigma_{A(B),\sigma}=U_{eff}\langle n_{A(B),\bar{\sigma}}\rangle$ which is independent of momentum and frequency. The kinetic energy term has corrections from the interactions due to the drive-induced correlated hopping, which gives $\epsilon_{k\sigma}=\epsilon_0(k)\big(1+t_{corr}(\langle n_{A\bar{\sigma}}-n_{B\bar{\sigma}}\rangle$\big). $\hat{G_\sigma}$ has poles at $E_{1(2),\sigma}(k)=\big[f_{A\sigma}(k)+f_{B\sigma}(k) \pm \sqrt{(f_{A,\sigma}(k)-f_{B,\sigma}(k))^2+4\epsilon_{k\sigma}^2}\big]/2$, where the subscripts $1$ and $2$ correspond to the conduction (+) and valence (-) bands, respectively. We solve numerically for the self-consistent values of the spin-dependent sublattice densities, $\langle n_{A(B)\sigma} \rangle$. \\
\noindent
\section {Results for the Hole-Doped System}
We employ the drive protocol in which the phase $\Phi$ is tuned with the drive amplitude and frequency being fixed.

Panels [a,b] of Fig.~\ref{ms_n} show the staggered magnetization $m_s=(m_{zB}-m_{zA})/2$ and the uniform magnetization $m_f=(m_{zA}+m_{zB})/2$, where $m_{z\alpha} = n_{\alpha\uparrow}-n_{\alpha\downarrow}$ with $\alpha=A,B$, for $n=0.95$ for $\Omega=15$ and $10$ with drive amplitude $V=2\Omega$. Starting with a weakly interacting paramagnetic metal ($U=1.0$, $n<1$), increasing $\Phi$ enhances $U_{eff}$ (as shown in Fig.[\ref{couplings}]) and triggers a first-order transition to a ferrimagnetic phase at an n-dependent threshold $\Phi_c$. This spin-symmetry breaking is marked by discontinuous jumps in both $m_s$ and $m_f$. As $\Phi$ increases towards $\pi/2$, $m_s$ continues to grow, while $m_f$ remains fixed at $n-1$.

\begin{figure}
\hskip-0.02in
\includegraphics[width=1.0\columnwidth,height=2.3in]{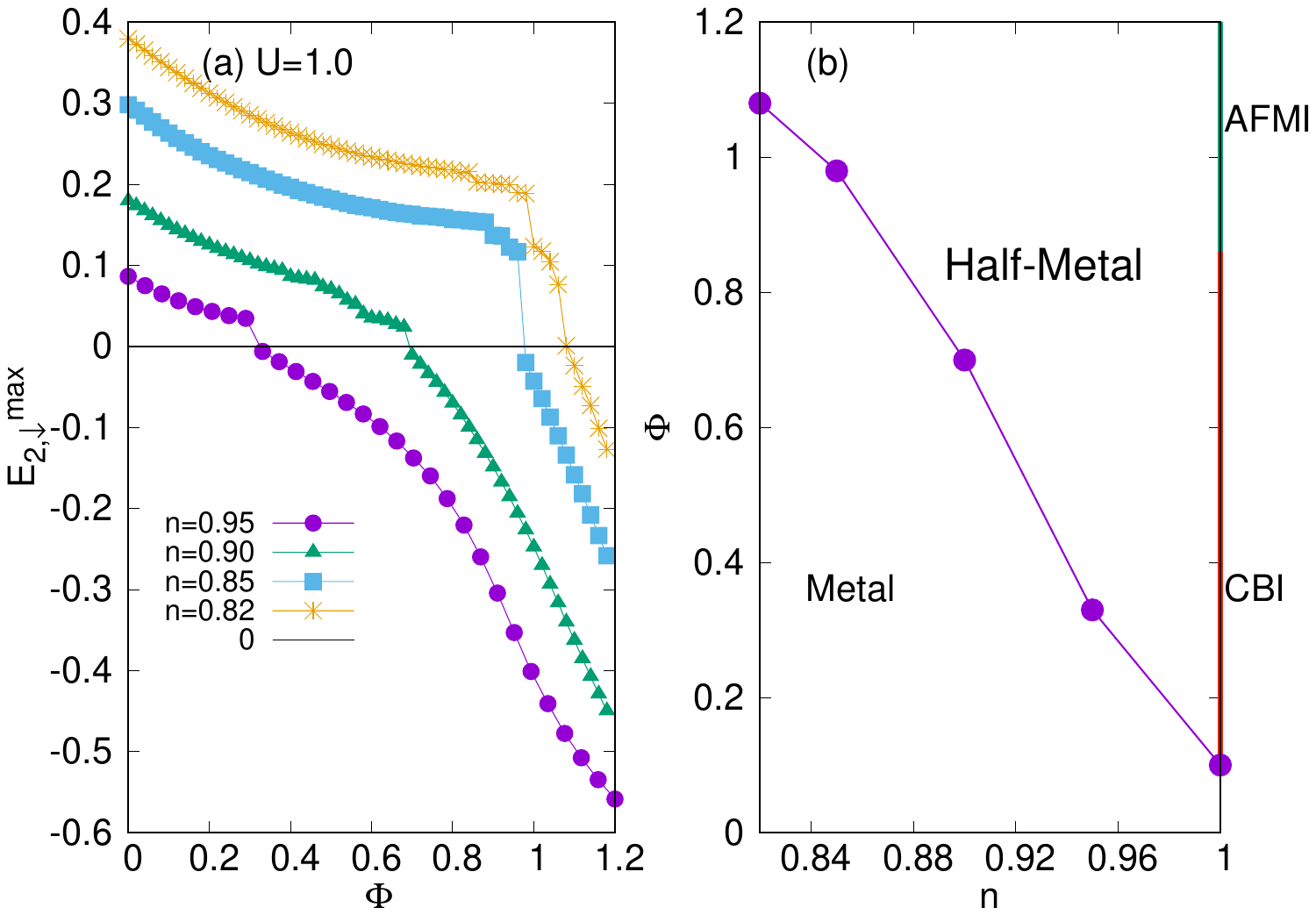}
\caption{Panel (a): The top of the down-spin valence band, $E_{2\downarrow}^{max}$, as a function of $\Phi$ for various values of $n$. Panel (b): Phase diagram in the $\Phi-n$ plane for the drive with $V=60$ and $\Omega=30$. }
\label{pd}
\end{figure}

The bottom panels (c,d) in Fig.~\ref{ms_n} show $E_{2\sigma}^{max}$, the top of  the valence bands (relative to $\mu_{eff}$) for the two spin orientations, as functions of $\Phi$. The conduction bands for both spins are above $\mu_{eff}$ for all $\Phi$. For $\Phi < \Phi_c$, when the system has spin degeneracy, $E_{2,\sigma}^{max}$ is positive, implying that the Fermi level lies inside the valence bands of both spin components, whence the system is a paramagnetic metal. In contrast, for $\Phi> \Phi_c$ $E_{2\uparrow}^{max}$ is positive whereas $E_{2\downarrow}^{max}$ is negative, implying that the up-spin electrons are conducting while the down-spin electrons completely fill their valence band and are insulating; hence, the system is a ferrimagnetic HM. Since the ferrimagnetic phase is a HM here for all $\Phi$, it is easy to understand why the uniform magnetization $m_f$ does not increase with $\Phi$. In the HM phase, since the down-spin valence band is fully occupied, all the holes are doped in the up-spin valence band. Thus,$n_\downarrow = 1/2$ while $n_\uparrow = n- 1/2$. Hence, $m_f = (n_\uparrow - n_\downarrow) = n-1$ independent of $U_{eff}$ and $\Phi$ for $\Phi> \Phi_c$. 

The rightmost panel in Fig.~\ref{ms_n} shows the phase diagram in $\Phi-\Omega$ plane for $n=0.95$ and $U=1.0$ when the drive amplitude is $V=2\Omega$. Here $\Phi_c$ is the value of $\Phi$ at which the half-metallic phase turns on. As the drive frequency increases, the ferrimagnetic half-metallic phase turns on at a smaller value of $\Phi$ and remains for a wider range of $\Phi$.  Fig.~\ref{ms_n2} shows the magnetizations and $E_{2\sigma}^{max}$ vs $\Phi$ for a much stronger drive with $\Omega=30$ and $V=60$ for $n=0.95$ and $n=0.85$. We have also calculated the single-particle spin- and sublattice- resolved density of states (DOS)(in units of $t_{eff}^{-1}$) $\rho_{\alpha,\sigma}(\omega) = -\frac{1}{\pi}\sum_k Im [G_{\alpha,\sigma}(k,\omega^{+})]$ which consistently shows the presence of half-metallic phase (Appendix-C). We emphasize that drive-induced higher-range hopping and correlated hopping are crucial for stabilizing a wide half-metallic phase, and without them, the system remains a paramagnetic metal as shown in~\cite{Bag}.

Across all studied densities, the HM phase emerges for $\Phi_c < \Phi \leq \pi/2$. The Fermi level is within the valence band for {\em both} spins for $\Phi < \Phi_c$, while the top of the down-spin valence band moves below the Fermi level for $\Phi > \Phi_c$ (cf. panel (a) of Fig.~\ref{pd}). $\Phi_c$ increases with decreasing $n$, resulting in a wider HM phase for densities closer to one (cf. panel (a) of Fig.~\ref{pd}). For, as $n$ reduces, a larger $U_{eff}$, and hence a larger $\Phi_c$, is required to turn on the magnetic order. As shown in Appendix-D, this trend remains the same even for the drives with lower drive frequency and amplitude as long as the drive parameters are much larger than the intrinsic energy scales of the undriven system.

While our results are presented in dimensionless units to maintain generality, in Table~\ref{table} we have mapped our key parameters to physical units to facilitate experimental realization, details of which are given in Appedix-A. 

\begin{table}
    \begin{tabular}{|c|c|c|c|}
    \hline
    System & $t_0 \sim U$   & $\Omega$    & $V$\\
    \hline
    Cold-atom & 100 Hz   & 1-3 KHz & (1.5-2.5)$\Omega$\\
    \hline
     TMDs   & 2 meV & 10-20 meV & $10^8$ W/cm$^2$\\
     \hline
    \end{tabular}
    \caption{Energy Scales for real experimental systems.}
    \label{table}
\end{table}

\begin{figure}[t]
    \includegraphics[width=3.6in]{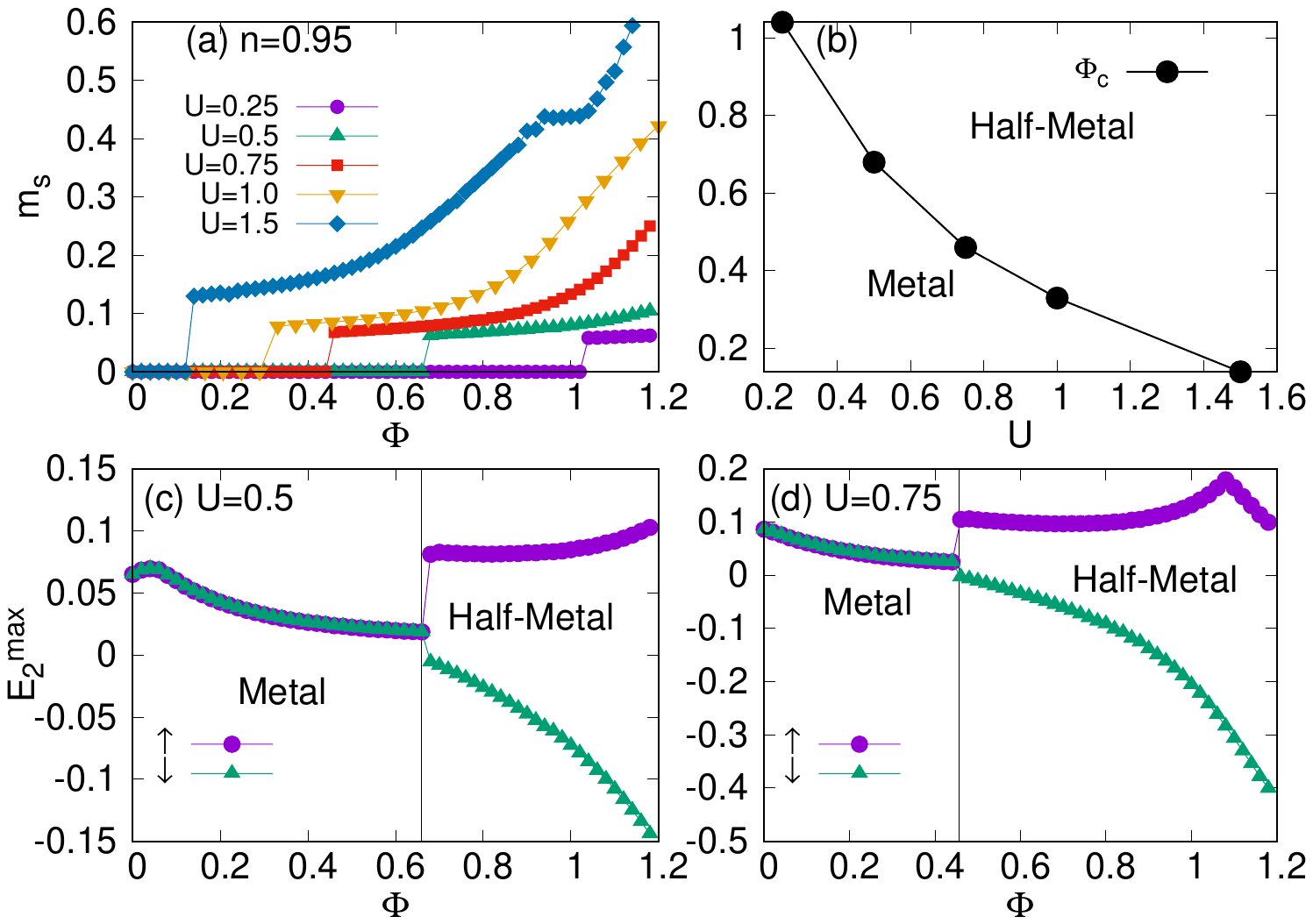}
    \caption{Panel (a): Staggered magnetization $m_s$ vs $\Phi$ for various values of $U$ for $n=0.95$. Panel (b): $\Phi_c$, where the magnetization and the HM turn on, as a function of $U$. Bottom panels: Maximum of $E_{2\sigma}(k)$, for both the spins, as functions of $\Phi$ for $U=0.5$ and $0.75$. The data shown are for the drive with amplitude $V=60$ and frequency $\Omega=30$.}
    \label{diffU}
\end{figure}

\noindent
{\bf{$U$-dependence of the Half-Metal Phase:}}
So far, we have presented results for the system with the value of the onsite interaction fixed at $U=1.0$ for various densities. To explore the effect of $U$ on the width of the drive-induced half-metal phase, we also analyzed the system for various values of $U$ for $n=0.95$. As $U$ increases, the magnetization turns on at smaller values of $\Phi$, thereby increasing the width of the half-metal phase.  This is because for larger $U$ values, $U_{eff}$ crosses the threshold value of interactions required to turn on magnetization at smaller values of the staggered phase $\Phi$ as seen from panel (a) of Fig.~\ref{diffU}. At $\Phi_c$ where the spin-symmetry is broken, the system transforms from a metal into a ferrimagnetic half-metal as shown in the lower panels of Fig.~\ref{diffU}. 
Due to the enhanced values of $m_s$ for larger values of $U$, the difference between $E_{2\uparrow}^{max}$ and $E_{2\downarrow}^{max}$ is also greater. This makes the half-metal phase more robust in systems with higher values of $U$. A plot of the threshold phase $\Phi_c$ vs U is shown in panel b of Fig.~\ref{diffU} for $n=0.95$. This shows how the width of the half-metal phase in $\Phi$ increases as $U$ increases. \\
\begin{figure}
\vspace{0.4cm}
\includegraphics[width=0.9\columnwidth,height=0.55\columnwidth]{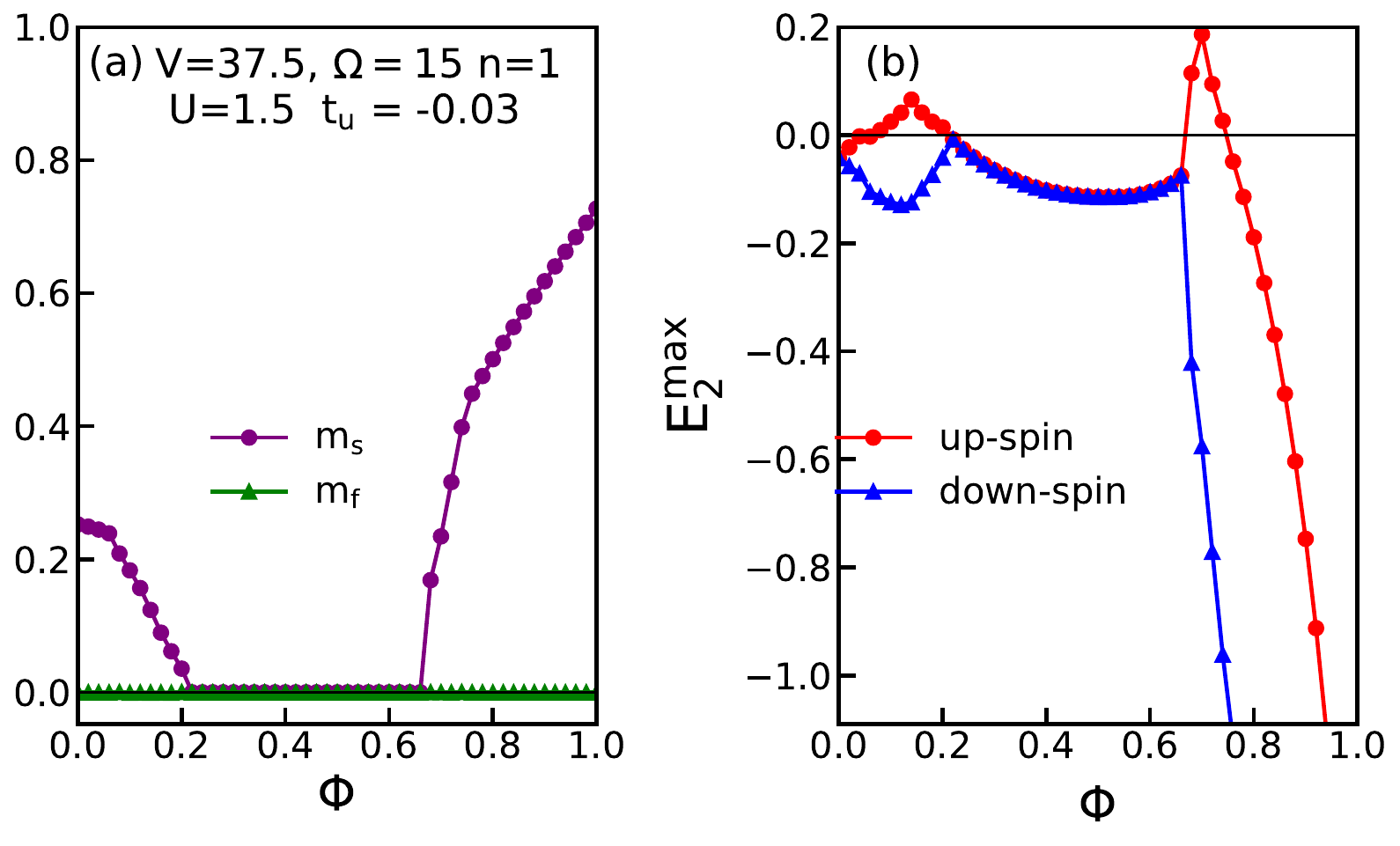}
\vspace{-0.3cm}
\caption{The left panel shows the staggered and uniform magnetization vs $\Phi$ for the half-filled system for $\Omega=15$ and $V=37.5$ for a tiny uniform 2nd neighbour hopping $t_u=-0.03$ in the undriven system. The right panel shows tops of the up-spin and down-spin valence bands, $E_{2\sigma}^{max}$ vs $\Phi$, as functions of $\Phi$.}
\label{tu}
\end{figure}
\section {Half-filled system}
The half-filled system is special and shows a very different phase diagram, with a reentrant phase transition. The undriven system with $U=1.0$ is an antiferromagnetic(AFM) insulator. With the drive, as $\Phi$ increases beyond $\Phi_1\approx 0.1$, the AFM order is lost as the enhanced $\Delta_{eff}$ and $t_{eff}$ transform the system into a paramagnetic band-insulator. Upon further increasing $\Phi$, the spin symmetry is broken again at $\Phi_2 \approx 0.86$ and the system is an AFM Mott insulator for all higher values of $\Phi$ up to $\pi/2$.
Near $\Phi_1$ and $\Phi_2$, the up-spin valence band nearly touches $\mu_{eff}$, allowing a tiny Zeeman field to readily induce a HM phase (see SM~\cite{SM}). Alternatively, introducing a tiny second-neighbor hopping $t_u$ stabilizes the half-filled HM phase. A very tiny $t_u$ also gets effectively enhanced due to the reduction in $t_{eff}$ in the periodically driven system, making the half-metal phase stable. As shown in Fig.~\ref{tu}, the system undergoes a series of transitions from AFM insulator to AFM half-metal to paramagnetic insulator and then again an AFM half-metal as $\Phi$ increases, although for a slightly larger value of $t_u$, the system remains in an AFM HM phase in the entire range of $\Phi$ as shown in Appendix-E. 

We note that all the calculations presented in this paper have been done for parameters such that $U_{eff} < 3.5$. Earlier work on the IHM using dynamical mean-field theory (DMFT)~\cite{IHM_AG1,IHM_Craco,Dagotto,IHM_Kampf,IHM_AG2,IHM_Hoang,Anwesha2,Bag} and determinant quantum Monte Carlo (DQMC) ~\cite{Scalettar1,Scalettar2} has shown that in such an intermediate interaction regime, there is a strong qualitative, and sometimes even quantitative, consistency between simple HF theory and the more sophisticated calculations like DMFT or DQMC. Thus, we believe that HF theory is sufficient to provide a reliable phase diagram of the periodically driven Hubbard model in the parameter regimes we have considered in this paper. However, as shown in Fig.~\ref{couplings}, there are parameter ranges for which the system can be in an extremely correlated regime with very high values of $U_{eff}$, which can lead to interesting consequences such as superconductivity~\cite{Anwesha2,Anwesha3}, an area for future exploration. 

\begin{figure}[t]
 \vspace{-0.5cm}
    \includegraphics[width=1.0\columnwidth,height=0.7\columnwidth]{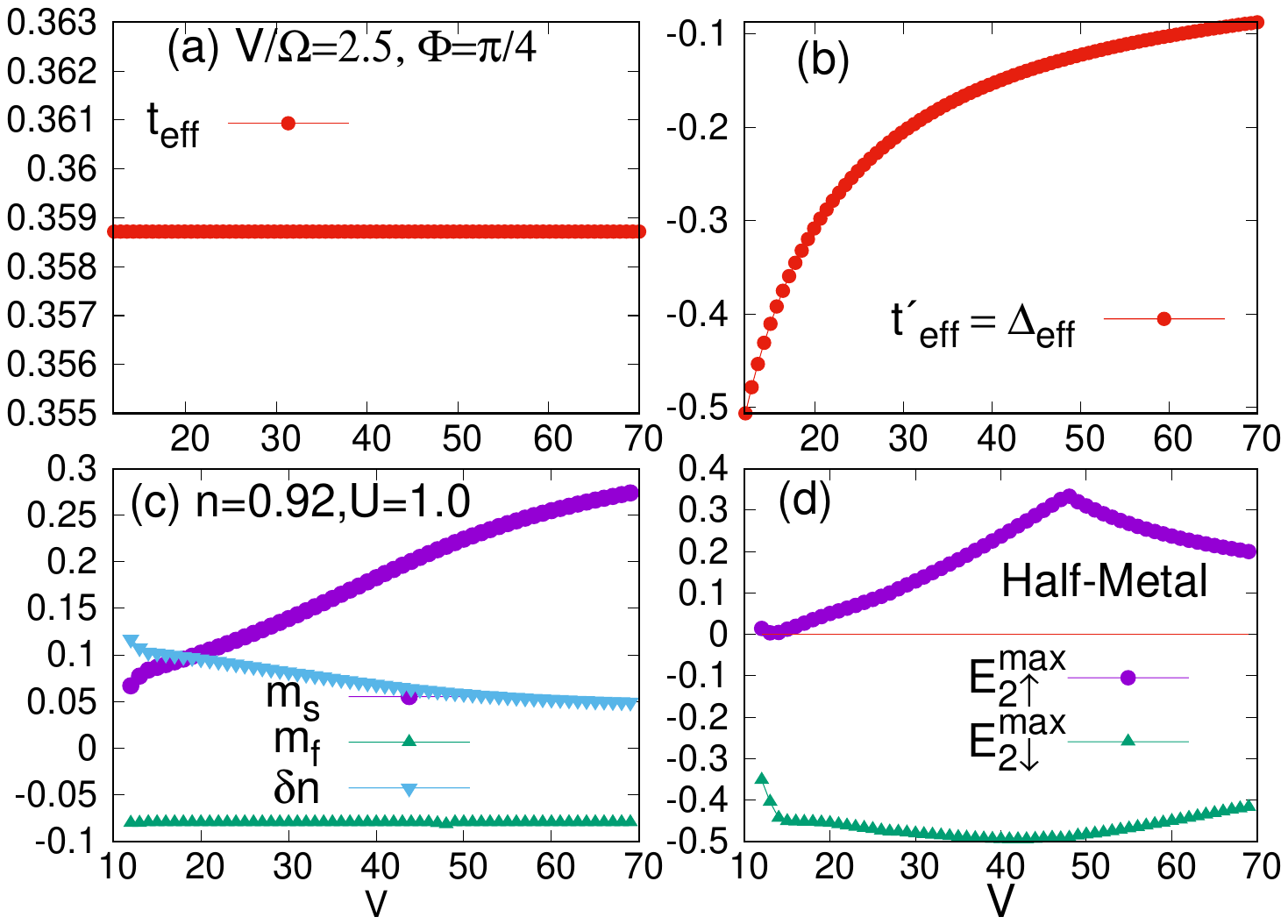}
    \vspace{-1.0cm}
    \caption{Results for the second driving protocol. Panel (a) and (b): The couplings $t_{eff}$ and $t^\prime_{eff}=\Delta_{eff}$ as functions of $V$ for $V/\Omega=2.5$ and $\Phi=\pi/4$. Panel (c): Self-consistent parameters as functions of $V$ for $n=0.92$ and $U=1.0$. Panel (d): $E_{2\sigma}^{max}$ as functions of $V$.}
    \label{phipi4}
\end{figure}

\noindent
{{\bf Alternate Driving Protocol:}}
We have also studied the system in an alternate drive protocol in which $\Phi$ is kept fixed at $\pi/4$ while the amplitude and frequency of the drive are tuned keeping the ratio $V/\Omega$ fixed at $2.5$, such that $t_{eff} (=0.3589)$ is small enough but the system is kept away from the DL point. As $V$ increases, the properties of the system are governed mainly by the drive-induced higher range hoppings and staggered potential because $U_{eff}$ in this protocol is fixed. With increasing $V$, $|t^\prime_{eff}|$ and $|\Delta_{eff}|$ decrease, whence $m_s$ is suppressed, while $\delta n$ is enhanced. $m_f$ stays fixed at $n-1$ for all $V$ values inside the half-metal phase. Interestingly, in this protocol, the system hosts a ferrimagnetic HM phase for a wide range of parameters. See Fig.~\ref{phipi4}.   

 \noindent
\section{Conclusions and Outlook} 
In conclusion, we have studied the periodically driven Hubbard model on a bipartite lattice in which the drive induces a staggered potential, staggered higher-order hoppings, and reduced first neighbor hopping, effectively equivalent to enhanced e-e interactions, leading to a ferrimagnetic half-metallic phase. We studied the system with the driving protocol, where the drive amplitude and frequency are fixed at values significantly higher than any energy scale in the system. As we turn on the drive in a weakly interacting hole-doped Hubbard model, a ferrimagnetic order sets in via a first-order transition at some threshold value of the staggered phase-angle $\Phi$ of the drive. The width of the half-metal phase along $\Phi$ is the largest for systems with densities closest to but below half-filling. For a fixed density, the width of the HM phase is larger for systems with larger values of the e-e interaction. As $\Phi$ increases, the half-filled system shows a re-entrant transition from an AFM insulator to a paramagnetic band insulator to an AFM Mott insulator. However, a tiny second neighbour hopping or a Zeeman field can induce half-metallicity even at half-filling in the presence of the drive. 

For high drive frequencies, i.e., in the prethermal regime, it should be possible to observe the proposed half-metal phase for times that are exponentially large in the drive amplitude and frequency. 
The drive parameters used here are in the off-resonant regime of the cold-atom experiments for periodically driven Fermi Hubbard~\cite{Esslinger1} and are not large enough to cause transition to higher bands in these systems. 
We believe that our work provides a new pathway for realizing a broad, robust half-metallic phase in the periodically driven weakly interacting Hubbard model, and it would be interesting to see these theoretical proposals getting tested in experiments. Our predictions are testable in various quantum simulator platforms, e.g., ultra-cold atoms on optical lattice~\cite{Bloch_Fermi_Hubbard, Bloch_RMP,Esslinger1,Esslinger2}, in Google Sycamore chip~\cite{Google_Fermi_Hubbard}, quantum dot arrays~\cite{Quantum_Dot_array_Fermi_Hubbard}, and artificial superlattice ~\cite{Cano,Chandni} 
and with improved technologies possibly even in real material systems like transition metal dichalcogenide heterostructures~\cite{TMD1,TMD2}.

Finally, generating stable, easily tunable phases of matter via periodic driving in the appropriate drive regime can herald a new way of designing quantum devices whose hardware would constitute stable, easily tunable quantum matter. 

\section{Acknowledgments}


AG would like to thank Krishnendu Sengupta for comments on the manuscript and U. Chandni for discussions on the experimental realization of the periodic drive in artificial superlattices. AG also acknowledges Mohit Randeria for discussions on related work. AD thanks Asmi Haldar, Anirban Das, and Sagnik Chaudhuri for various useful discussions on the stability of many-body systems under a strong periodic drive. HRK acknowledges support from the Indian National Science Academy under grant no. INSA/SP/SS/2023/. A.G. acknowledges support from
the Anusandhan National Research Foundation, Government of India (Grant No. ANRF/ARGM/2025/000622/TS).

\section{ Appendix-A: Implementation of the periodic drive}
In the main text we briefly described the possible experimental realization of the periodic drive explored in this work in cold-atom experiments. Here we provide further details. 
An ideal platform to implement periodic drive of the Fermi Hubbard model is the ultra-cold atomic systems. In the context of cold-atom systems, two sublattice periodic drive can be implemented by using two different lasers to generate A and B sublattices which are rotated at $45^\circ$ and $135^\circ$ angles, respectively, with respect to the x-axis. Two mirrors are used to generate the standing waves corresponding to two sublattices. In order to periodically shake the two sublattices, one needs to use piezoelectric actuators attached to the mirrors generating the separate standing waves corresponding to the two sublattices which effectively periodically shakes each sublattice independently.  
In order to have the required phase difference $\Phi$ between the two sublattice drives, we need to turn on the drive on two subllatices at different times. The time lag will mimic the effect of the phase difference in our drive. For a drive with frequency $\Omega$ (considering $\bar{h}=1$), a phase difference of $\Phi$ can be obtained by having a time lag of $\Delta t =2\Phi/\Omega$.

{\bf{Parameter Regime for the periodic drive:}} The high-frequency periodic drives must be carefully tuned to avoid excitations to higher bands. To ensure the stability of the proposed half-metallic phase, the system should operate in the regime:$$\Omega \gg \{U, t_0\} \quad \text{and} \quad \Omega \ll E_g$$ where $E_g$ is the energy gap to the higher band. Though we have worked with a single-band Hubbard model, our calculations explicitly account for these constraints in possible cold-atom experimental set ups. The results shown in this work are obtained for the periodic drive with $5<\Omega/t_0 \le 30$ and $5<\Omega/U \le 30$ and the drive amplitude is $1.875 < V/\Omega \le 2.5$. The gap between the s and p bands, $E_g$, in the optical lattice system is of the order of 5 to 20 times the recoil energy, which is equivalent to 20kHz$< E_g <35$ kHz. In recent simulations of periodically driven Fermi Hubbard model~\cite{Esslinger1}, hoppings $(t_x,t_y,t_z) \sim (200,40,40)$Hz and $U = 500Hz$ and the periodic drive with frequency $\Omega=7.25$kHz is used with an amplitude of $1.69\Omega$. In the above mentioned experiment, a periodic drive with $\Omega \sim 15 U \sim 180 t_y/t_z$, does not cause excitations to higher band which is clear from a good consistency between experimentally measured values of double occupancies and their comparison with those obtained by solving the effective Hamiltonian obtained by Magnus expansion of the single band Hubbard model in the presence of the drive. Based on these details, we have worked with a periodic drive with $5t_0 < \Omega \le 30t_0$ and $5U< \Omega \le 30U$, which is in the appropriate regime for experimental realization of our findings in ultracold atom systems. 
\\\\
{\bf Realization in real material systems}: 
The best platform in real material systems where the effect of periodic drive may be realized in future are twisted transition metal dichalogenide (TMD) heterostructures like $MoTe_2/WSe_2$ heterobilayers~\cite{TMD1,Li}, $MoSe_2/WSe_2$ heterostructures ~\cite{Zhou} etc. In the $MoTe_2/WSe_2$ heterobilayers, depending on the twist angle, the global band gap $E_g$ can vary from 0.8eV $< E_g < 2.5$eV, the separation between the 1st Moire miniband and the 2nd miniband is from $(100-150)$ meV while the band width $W_b$ of the Moire miniband is generally of the order of a few  meV with  5meV $< W_b < $15meV . Similarly in twisted $WS_2/WSe_2$, the gap between Moire bands is of the order of 60meV while $W_b \sim (2-5)$meV which implies a hopping amplitude $t$ of less than a meV.  

In such cases, one can drive the system with a drive frequency of $ 2W_b< \Omega <10W_b$ still keeping $\Omega < E_g$. Due to the twist angle, there is lot of freedom in tuning the band gap and the band width of Moire bands though the sublattice structure of the drive may not be easy to implement in these materials without significant improvement of technology. Another platform where our proposed half metal can be realized
are artificial superlattices which are able to generate Moire potentials using an engineered
periodic array of gates~\cite{Cano,Chandni}. It should be possible to drive the two sublattices differently in these
system with time dependent gate voltages.
\\\\
\noindent

\section{Appendix-B: Details of Floquet Hamiltonian}
In this section, we provide details of the Floquet Hamiltonian, $H_F$, described in Eqn.[4] of the main text. 
As mentioned in the main text, the time dependence in the Hamiltonian in Eqn.[4] can be transformed into an effective time-dependent hopping by going to a rotating frame $H_{rot} = S^\dagger(t) [H_{hub} +H_{drive} ]S(t)-iS^\dagger(t)\partial_t S(t)$, where $S(t)=\exp\big[-i\sum_j F_j(t) n_j\big]$ with $F_j(t)=\frac{V}{2}\int_0^t \sin(\Omega t \mp \Phi)$ for $j\in A(B)$ respectively~\cite{Polkovnikov}.  Note that the Hamiltonian in the rotating frame has acquired an extra Galilean term due to the fact that the transformation $S(t)$ is time dependent. 

The kinetic energy in the rotating frame is
\begin{equation}
H_{hop}(t)=-t_0g(t)\sum_{\langle ij\rangle,\sigma}c^\dagger_{iA\sigma}c_{jB\sigma}+ h.c.
\end{equation}
with $g(t)=\exp(i\frac{V}{\Omega}\sin(\Omega t)\sin(\Phi))$ while the $H_{drive}$ gets eliminated in the rotated frame . 

The other terms in the Hamiltonian $H_{hub}$ remain unchanged under rotation. Using the Jacobi-Anger identity, one can write the time-dependent hopping term in the form of a series as shown below:
\begin{equation}
H_{hop}(t)=-t_0\sum_{l=-\infty}^{\infty} \mathcal{J}_l(\frac{V}{\Omega}\sin(\Phi)) e^{il\Omega t} \sum_{\langle ij\rangle,\sigma}c^\dagger_{i\sigma}c_{j\sigma}+ h.c.
\end{equation}
Here, $\mathcal{J}_l(\frac{V}{\Omega}\sin(\Phi))$ is the $l^{th}$ Bessel function of the first kind.
The stroboscopic evolution operator for the time dependent Hamiltonian is defined as 
\begin{equation}
    U(T,0)=\mathcal{T}_t \exp\big(-\frac{i}{\hbar}\int_0^T d\tilde{t}~H[\tilde{t}]~\big)\equiv \exp\big[-\frac{i}{\hbar} H_F~T\big]
    \label{HF_defn}
\end{equation}
Taking log on both sides of Eqn.~\ref{HF_defn}, one gets a series expansion for the Floquet Hamiltonian $ H_F =\sum_{n=0}^{\infty} H_F^{(n)}$.
Keeping terms up to order $1/\Omega$ in the Magnus expansion results in the following form of the Floquet Hamiltonian $H_F = H_F^{(0)}+H_F^{(1)}$ with 
\begin{equation}
    H_F^{(0)} = \frac{1}{T}\int_0^{T} H(t) dt
    \end{equation}
    and 
    \begin{equation}    
    \begin{split}
    H_F^{(1)} = \frac{1}{2T i \hbar}\int_{0}^{T} dt_1 \int_{0}^{t_1} dt_2 [H(t_1),H(t_2)] \\
    = \frac{1}{4\pi\Omega i\hbar} \int_0^{2\pi} d \tau_1 \int_0^{\tau_1} d\tau_2 [H(\tau_1),H(\tau_2)]
    \end{split}
    \end{equation}
    
    with $\tau=\Omega t$.
Using the Hamiltonian in the rotated frame $H(t)=H_{hopp}(t)+U\sum_i n_{i\uparrow}n_{i\downarrow}-\mu\sum_i n_i$, we obtain the following form of the Floquet Hamiltonian, 
\begin{widetext}
\begin{equation}
\begin{split}
H_F^{[0]} = -t_0 \sum_{\langle ij \rangle, \sigma} \left( \mathcal{J}_0(\frac{V}{\Omega}sin(\Phi)) c_{i,\sigma}^\dagger c_{j,\sigma} + h.c \right) + U \sum_j n_{j \uparrow} n_{j \downarrow} -\mu\sum_i n_i, \\
H_F^{[1]}  = \frac{t^2}{4\pi i\hbar \Omega} \int_0^{2\pi} dt_1 \int_0^{t_1} dt_2 \sum_{\langle ij\rangle,\sigma} 
\left[
\big( g(t_1) g^*(t_2) - g^*(t_1) g(t_2) \big) 
\big( n_{i \sigma}^A - n_{j, \sigma}^B \big)
\right] \\
+ \frac{2t^2}{4 \pi i\hbar \Omega} \int_0^{2\pi} dt_1 \int_0^{t_1} dt_2 \sum_{(ij),\sigma}
\left[
\big( g(t_1) g^*(t_2) - g^*(t_1) g(t_2) \big) 
\big( c_{iA,\sigma}^\dagger c_{jA,\sigma}
- c_{iB, \sigma}^\dagger c_{jB, \sigma} \big) 
\right] +h.c.\\
+ \frac{t^2}{4 \pi i\hbar \Omega} \int_0^{2\pi} dt_1 \int_0^{t_1} dt_2 \sum_{((ij)),\sigma}
\left[
\big( g(t_1) g^*(t_2) - g^*(t_1) g(t_2) \big) 
\big( c_{iA,\sigma}^\dagger c_{jA,\sigma}
- c_{iB, \sigma}^\dagger c_{jB, \sigma} \big) 
\right] +h.c.\\
+ \frac{tU}{4\pi i\hbar \Omega} \int_0^{2\pi} dt_1 \int_0^{t_1} dt_2 \sum_{\langle ij\rangle}^z \sum_\sigma 
\left[
\big( g(t_1) - g(t_2)\big) \big(n_{i \sigma}^A -n_{j, \sigma}^B \big)c_{iA \bar{\sigma}}^\dagger c_{jB, \bar{\sigma}} + h.c.
\right].
\end{split}
\label{HF1}
\end{equation}
\end{widetext}

\section{Appendix-C: Single Particle Density of States}
 \begin{figure}[t]
\hskip-0.1in
    \includegraphics[width=0.75\columnwidth]{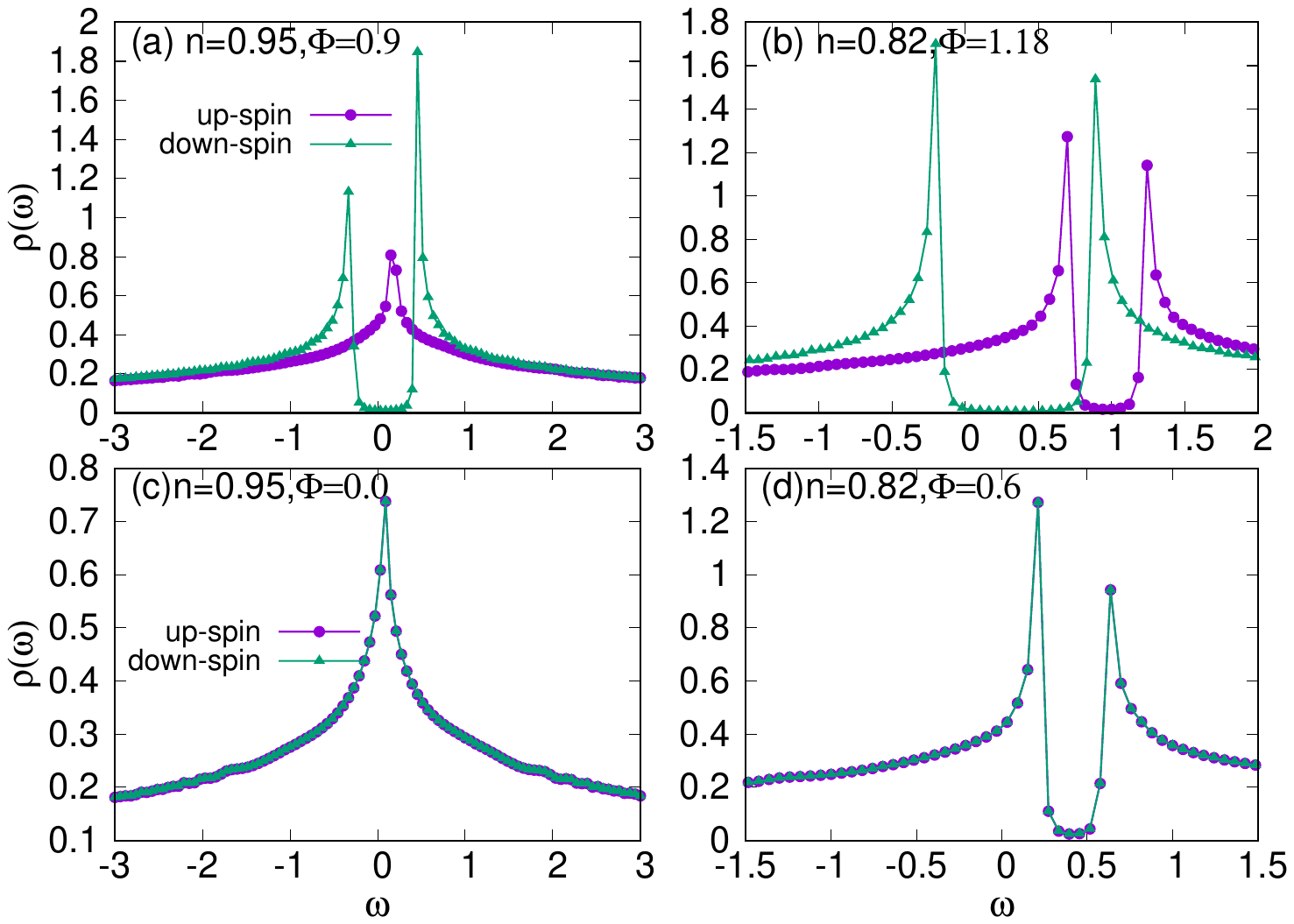}

    \caption{Single-particle spin-resolved density of states $\rho_\sigma(\omega)$ vs $\omega$ for $U=1.0$ and two different densities. Bottom Panels: For $\Phi < \Phi_c$, $\rho_\sigma(\omega=0)$ is finite for both the spin-components. 
    Top Panels: For $\Phi > \Phi_c$, $\rho_\uparrow(\omega=0)$ is finite while $\rho_{\downarrow}(\omega=0) =0$. This is the HM phase.
    The data shown are for the drive with $V=60$, $\Omega=30$.}
    \label{dos}
\end{figure} 

We have also calculated the single-particle spin- and sublattice- resolved density of states (DOS)(in units of $t_{eff}^{-1}$) $\rho_{\alpha,\sigma}(\omega) = -\frac{1}{\pi}\sum_k Im [G_{\alpha,\sigma}(k,\omega^{+})]$. 
Fig.~\ref{dos} shows $\rho_\sigma(\omega)$, the sum of the two sublattice DOS, for both spins as functions of $\omega$. For $\Phi < \Phi_c$, where the system has spin degeneracy, $\rho_{\uparrow}(\omega)=\rho_{\downarrow}(\omega)$, and $\rho_\sigma(\omega=0)$ is finite for both spins. This is the paramagnetic metallic phase (cf. panels (c) and (d) of Fig.\ref{dos}). For $\Phi> \Phi_c$, as in panels (a) and (b) of Fig.~\ref{dos}, the Fermi level is inside the valence band only for the up-spin; thus $\rho_{\uparrow}(\omega=0) \neq 0$ while $\rho_{\downarrow}(\omega=0) = 0$. This is the drive-induced HM phase for which the up-spin band is conducting and the down-spin band is insulating. 

\section{Appendix-D: Density dependence of the HM phase for lower frequency drive}
In the main paper we have shown the phase diagram in $\Phi-n$ plane for $n <1$ for the drive frequency $\Omega=30$ and drive amplitude $V=60$. This density dependence of the half-metal phase is not restricted to the large frequency and amplitude drive shown in the main paper but in fact persists for weaker drives as well as long as $\Omega,V \gg U,t_0$. Here we are showing the similar results for a drive with $\Omega=10$ and $V=20$. As $n$ increases, the half-metal phase turns on at a smaller value of $\Phi$ which itself increases as one goes to lower drive frequencies.
\begin{figure}
  \begin{center}
  \includegraphics[width=2.3in,angle=0]{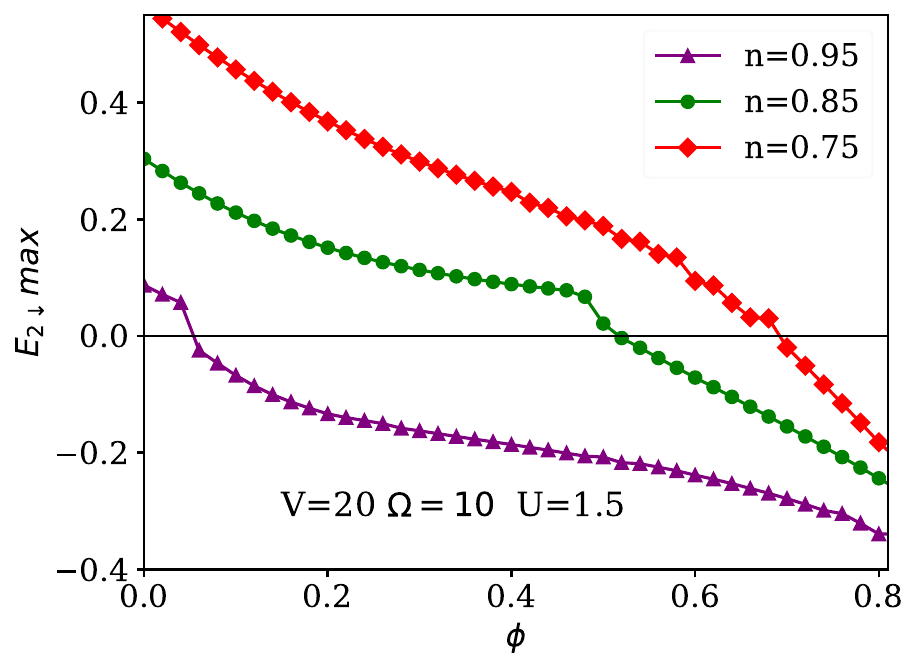}
          \caption{Top of the valence band for down spins as functions of $\Phi$ for various values of densities. Data shown is for $\Omega=10, V=20$ and initial $U=1.5$.}
          \label{lowdrive4}
          \end{center}
           \end{figure}

\section{Appendix-E: Self-consistent parameters for the half-filled system}
In the main text, we presented the results for the hole-doped system. Here, we do so for the periodically driven Hubbard model at half-filling. Note that the second and third neighbour hopping terms do not break the particle-hole symmetry of the half-filled system. Hence, the chemical potential for the half-filled system is pinned at $U/2$.  
\begin{figure}
\vspace{0.7cm}
  \includegraphics[width=0.95\columnwidth,height=0.6\columnwidth]{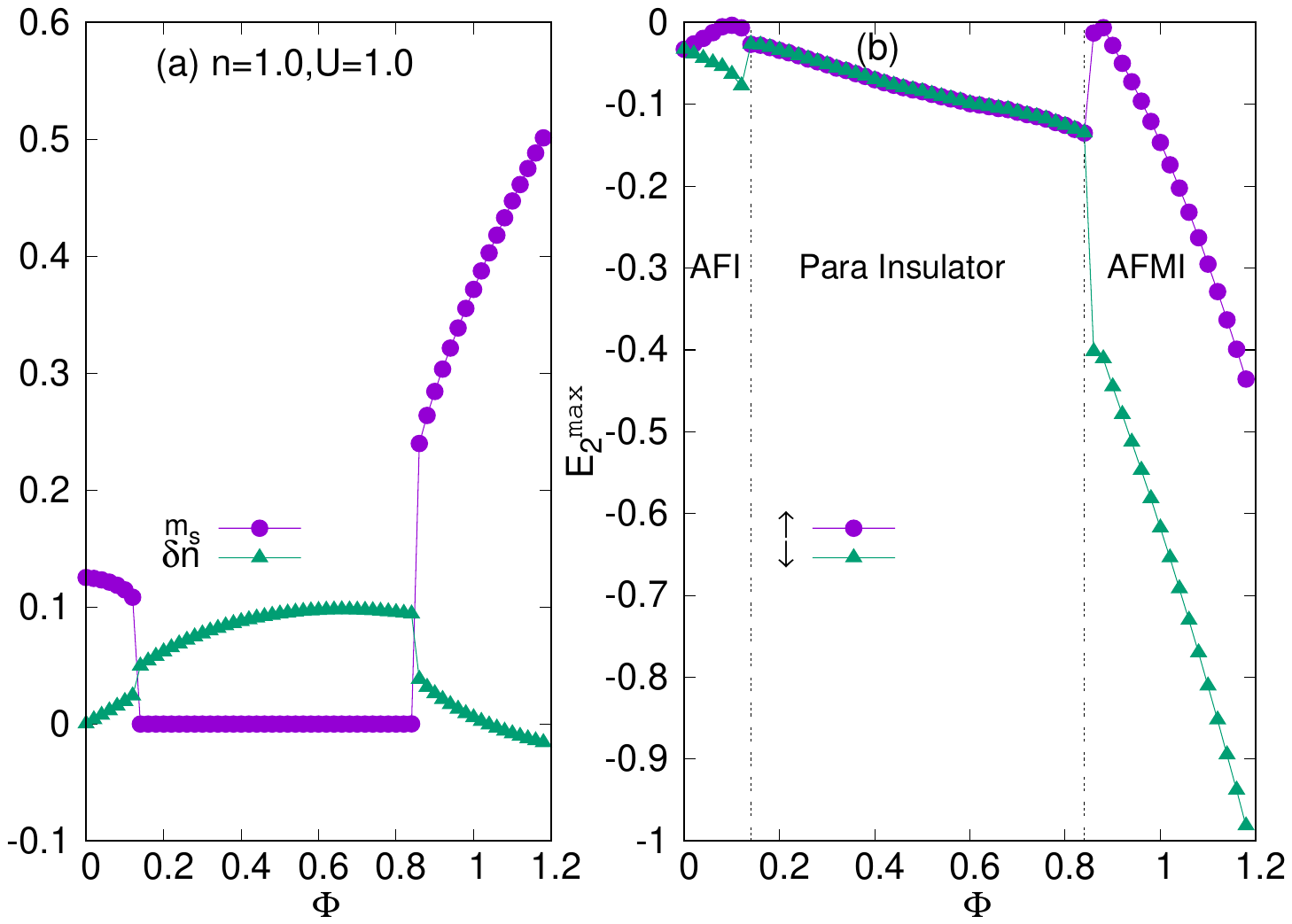}
  \vskip-0.2cm
  \includegraphics[width=0.85\columnwidth,height=0.5\columnwidth]{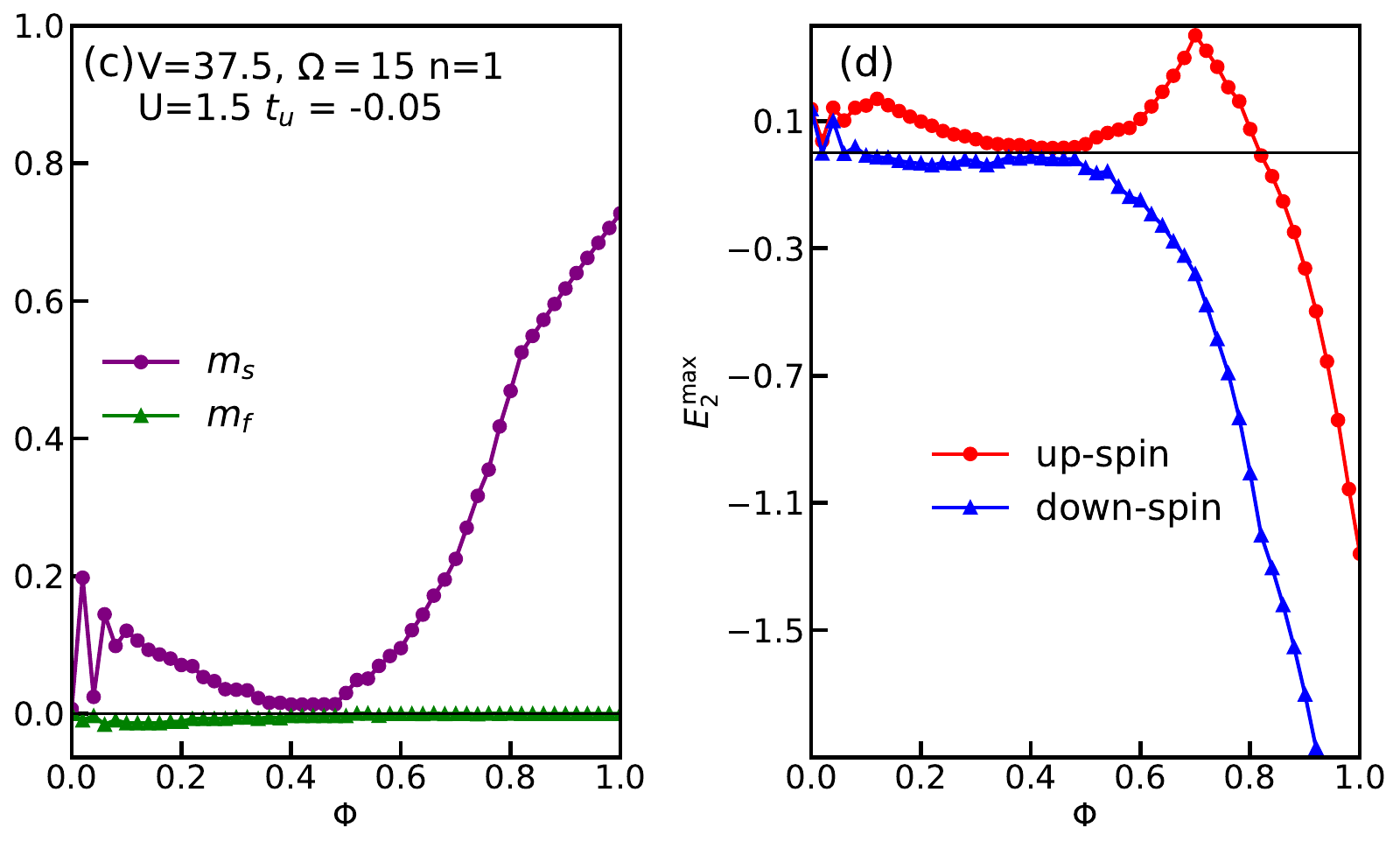}
   \caption{Panel [a,b]: The staggered magnetization $m_s$ and the sub-lattice density difference $\delta n=(n_A-n_B)/2$ as functions of $\Phi$ for $n=1$. Panel [b] shows the maximum value of $E_{2\sigma}(k)$ for the two spins as functions of $\Phi$. The data shown are for the drive with amplitude $V = 60$ and frequency $\Omega = 30$.
   Panels [c,d]: Results at half-filling in the presence of a small next neighbour hopping $t_u=-0.05t_0$ on diagonal bonds in the square lattice in the Hamiltonian of Eqn(1) in the main paper. Even such a  small second neighbour hopping can induce half-metallicity at half-filling in driven system.}
\label{n1}
\end{figure}

Fig.~\ref{n1} shows the self-consistent parameters for a half-filled driven system starting from $U=1$. The undriven system has antiferromagnetic order and a spectral gap equal to $Um_s$ (for arbitrarily small $U$ because of the nested Fermi surface of bipartite lattices). As the drive is turned on and $\Phi$ is tuned, due to the drive-induced staggered potential and higher range hoppings,  the staggered magnetization decreases and goes to zero via a first order transition at $\Phi_1$($\approx 0.14$ for the parameters in Fig.~\ref{n1}). As $\Phi$ increases, and along with it $U_{eff}$ also increases overcoming the effect of staggered hopping and staggered potential, at $\Phi_2$ ($\approx 0.86$ for the parameters in Fig.~\ref{n1}) the anti-ferromagnetic order turns on again via another first order phase transition.  The density difference $\delta n$ increases with $\Phi$ with a jump across $\Phi_1$, but reaches a flattish maximum for $\Phi$ a bit less than $\Phi_2$ and then decreases with another jump across $\Phi_2$. The decrease in $\delta n$ for the large values of $\Phi$ is due to the dominating effect of enhanced $U_{eff}$ in the system.

Panel (b) in Fig.~\ref{n1} shows the tops of the valence bands for both spins as functions of $\Phi$. The undriven system is an antiferromagnetic insulator with an equal spectral gap in both spin bands. For $0.08<\Phi<0.12$, the spin-up valence band almost touches the Fermi-level ($|E_{2 \uparrow}^{max}| \leq 0.003$) while the down-spin band is below the Fermi-level and gapped. Thus, just before $\Phi_1$ the system can transform into an antiferromagnetic half-metal with a tiny Zeeman field. As $\Phi$ increases further, both spin bands remain gapped with equal gap and the system is a paramagnetic band insulator with a finite density difference between two sub-lattices. 
For $\Phi> \Phi_2$, with the onset of antiferromagnetic order, the system enters into an antiferromagnetic Mott insulator in which the gap increases with the increase in $U_{eff}$ but the spectral gaps in the up and down spin bands are unequal. Just after the AFM order turns on at $\Phi_2$, $|E_{2\uparrow}^{max}|<0.005$ around $\Phi\sim 0.88$ and a tiny Zeeman field can make this system a half-metal by pulling the chemical potential inside the valence band of the up-spin channel.

We would like to emphasize that unequal spectral gaps for the spin up and down bands at half-filling are crucial because when the system is doped below half-filling, the Fermi level moves below the top of the up-spin valence band while still remaining above the down-spin valence band. This results in the half-metallic phase for $n<1$ in this periodically driven Hubbard model, as we have discussed in detail in the main paper.

At half-filling, turning on a tiny 2nd neighbour hopping in the undriven system can stabilize the half-metal phase as shown in the main paper for like $t_u=-0.03$ where the system undergoes a re-enterant phase transition as $\Phi$ increases. But at a slightly larger value of $t_u=-0.05$, the system exhibits the AFM half-metallic phase for the entire range of $\Phi$ as shown in Fig.~\ref{n1}. This is in complete contrast to the occurrence of a HM phase at half-filling in the static IHM for a much larger threshold value of the second neighbor hopping~\cite{Bag}.

\bibliography{HM_drive.bib}

\end{document}